\documentclass[french,english,aps,manuscript,preprintnumbers,showpacs,showkeys]{revtex4-1}
\usepackage[T1]{fontenc}
\usepackage[latin9]{inputenc}
\usepackage{color}
\usepackage{amsmath}
\usepackage{amssymb}
\usepackage{esint}

\makeatletter
\@ifundefined{textcolor}{}
{%
 \definecolor{BLACK}{gray}{0}
 \definecolor{WHITE}{gray}{1}
 \definecolor{RED}{rgb}{1,0,0}
 \definecolor{GREEN}{rgb}{0,1,0}
 \definecolor{BLUE}{rgb}{0,0,1}
 \definecolor{CYAN}{cmyk}{1,0,0,0}
 \definecolor{MAGENTA}{cmyk}{0,1,0,0}
 \definecolor{YELLOW}{cmyk}{0,0,1,0}
}

\makeatother

\usepackage{babel}
\makeatletter
\addto\extrasfrench{%
   \providecommand{\fg}{\ifdim\lastskip>\z@\unskip\fi~\frqq}%
}

\makeatother
\begin{document}

\title{Exact solutions of the $\left(2+1\right)$-dimensional Dirac oscillator
under a magnetic field in the presence of a minimal length in the
noncommutative phase-space}

\author{Abdelmalek Boumali}

\email{boumali.abdelmalek@gmail.com}

\affiliation{Laboratoire de Physique Appliquée et Théorique, \\
 Université de Tébessa, 12000, W. Tébessa, Algeria.}

\author{Hassan Hassanabadi}

\email{h.hasanabadi@shahroodu.ac.ir}

\affiliation{Department of Physics, Shahrood University of Technology,\\
 Shahrood, Iran P.O. Box 3619995161-316 Shahrood Iran.}
\begin{abstract}
We consider a two-dimensional Dirac oscillator in the presence of
magnetic field in noncommutative phase space in the framework of relativistic
quantum mechanics with minimal length. The problem in question is
identified with a Poschl-Teller potential. The eigenvalues are found
and the corresponding wave functions are calculated in terms of hypergeometric
functions. 
\end{abstract}



\maketitle

\section{introduction}

The Dirac relativistic oscillator is an important potential both for
theory and application. It was for the first time studied by Ito et
al\cite{1}. They considered a Dirac equation in which the momentum
$\vec{p}$ is replaced by $\vec{p}-im\beta\omega\vec{r}$, with $\vec{r}$
being the position vector, $m$ the mass of particle, and $\omega$
the frequency of the oscillator. The interest in the problem was revived
by Moshinsky and Szczepaniak \cite{2}, who gave it the name of Dirac
oscillator (DO) because, in the non-relativistic limit, it becomes
a harmonic oscillator with a very strong spin-orbit coupling term.
Physically, it can be shown that the (DO) interaction is a physical
system, which can be interpreted as the interaction of the anomalous
magnetic moment with a linear electric field \cite{3,4}. The electromagnetic
potential associated with the DO has been found by Benitez et al \cite{5}.
The Dirac oscillator has attracted a lot of interest both because
it provides one of the examples of the Dirac's equation exact solvability
and because of its numerous physical applications (see \cite{6} and
references therein). Recently, Franco-Villafane et al \cite{7} exposed
the proposal of the first experimental microwave realization of the
one-dimensional DO. The experiment relies on a relation of the DO
to a corresponding tight-binding system. The experimental results
obtained, concerning the spectrum of the one-dimensional DO with and
without the mass term, are in good agreement with those obtained in
the theory. In addition, Quimbay et al \cite{8,9} show that the Dirac
oscillator can describe a naturally occurring physical system. Specifically,
the case of a two-dimensional Dirac oscillator can be used to describe
the dynamics of the charge carriers in graphene, and hence its electronic
properties. Also, the exact mapping of the DO in the presence of a
magnetic field with a quantum optics leads to regarding the DO as
a theory of an open quantum systems coupled to a thermal bath ( see
Ref \cite{6} and references therein).

The unification between the general theory of relativity and the quantum
mechanics is one of the most important problems in theoretical physics.
This unification predicts the existence of a minimal measurable length
on the order of the Planck length. All approaches of quantum gravity
show the idea that near the Planck scale, the standard Heisenberg
uncertainty principle should be reformulated. The minimal length uncertainty
relation has appeared in the context of the string theory, where it
is a consequence of the fact that the string cannot probe distances
smaller than the string scale $\hbar\sqrt{\beta}$, where $\beta$
is a small positive parameter called the deformation parameter. This
minimal length can be introduced as an additional uncertainty in position
measurement, so that the usual canonical commutation relation between
position and momentum operators becomes$\left[\hat{x},\hat{p}\right]=i\hbar\left(1+\beta p^{2}\right)$.
This commutation relation leads to the standard Heisenberg uncertainty
relation $\triangle\hat{x}\triangle\hat{p}\geq i\hbar\left(1+\beta\left(\triangle p\right)^{2}\right)$,
which clearly implies the existence of a non-zero minimal length $\triangle x_{\mbox{min}}=\hbar\sqrt{\beta}$.
This modification of the uncertainty relation is usually termed the
generalized uncertainty principle (GUP)or the minimal length uncertainty
principle\cite{10,11,12,13}. Investigating the influence of the minimal
length assumption on the energy spectrum of quantum systems has become
an interesting issue primarily for two reasons. First, this may help
to set some upper bounds on the value of the minimal length. In this
context, we can cite some studies of the hydrogen atom and a two dimensional
Dirac equation in an external magnetic field. Moreover, the classical
limit has also provided some interesting insights into some cosmological
problems. Second, it has been argued that quantum mechanics with a
minimal length may also be useful to describe non-point-like particles,
such as quasi- particles and various collective excitations in solids,
or composite particles (see Ref \cite{14} and references therein)

Nowadays, the reconsideration of the relativistic quantum mechanics
in the presence of a minimal measurable length have been studied extensively.
In this context, many papers were published where a different quantum
system in space with Heisenberg algebra was studied. They are: the
Abelian Higgs model \cite{15}, the thermostatics with minimal length
\cite{16}, the one-dimensional Hydrogen atom \cite{17}, the casimir
effect in minimal length theories \cite{18}, the effect of minimal
lengths on electron magnetism \cite{19}, the Dirac oscillator in
one and three dimensions \cite{20,21,22,23,24}, the solutions of
a two-dimensional Dirac equation in presence of an external magnetic
field \cite{25}, the noncommutative phase space Schrödinger equation\cite{26},
Schrödinger equation with Harmonic potential in the presence of a
Magnetic Field \cite{27}.

The study of non-commutative spaces and their implications in physics
is an extremely active area of research. It has been argued in various
instances that non-commutativity should be considered as a fundamental
feature of space-time at the Planck scale. On the other side, the
study of quantum systems in a non-commutative (NC) space has been
the subject of much interest in last years, assuming that non-commutativity
may be, in fact, a result of quantum gravity effects. In these studies,
some attention has been given to the models of non-commutative quantum
mechanics (NCQM). The interest in this approach lies on the fact that
NCQM is a fruitful theoretical laboratory where we can get some insight
on the consequences of non-commutativity in field theory by using
standard calculation techniques of quantum mechanics. Various non-commutative
field theory models have been discussed as well as many extensions
of quantum mechanics. Of particular interest is the so-called phase
space non-commutativity which has been investigated in the context
of quantum cosmology, black holes physics and the singularity problem.
This specific formulation is necessary to implement the Bose-Einstein
statistics in the context of NCQM (see Refs. \cite{28,29} and references
therein).

The purpose of this work is to investigate the formulation of a two-dimensional
Dirac oscillator in the presence of a magnetic field by solving fundamental
equations in the framework of relativistic quantum mechanics with
minimal length in the non-commutative phase space. To do this we first
mapped the problem in question into a commutative space by using an
appropriate transformations. Then, we solved it in the presence of
a minimal length. We would like mentioned here that the origin of
relativistic Landau problem and the Dirac oscillator is entirely different
in the former case the magnetic field is introduced via minimal coupling
while in the latter case the interaction is introduced via non-minimal
coupling and can be viewed as anomalous magnetic interaction \cite{30,31}.

The paper is organized as follows. In section II, we solve the Dirac
oscillator in the presence of magnetic field in noncommutative phase
space. Then, in section III, we study this problem in the framework
of relativistic quantum mechanics with minimal length. Finally, in
section IV, we present the conclusion.

\section{The solutions in Noncommutative Phase Space}

To begin with we note that the non-commutative phase space is characterized
by the fact that their coordinate operators satisfy the equation \cite{28,29}
\begin{equation}
\left[x_{i}^{\left(NC\right)},x_{j}^{\left(NC\right)}\right]=i\tilde{\theta}_{ij},\,\left[p_{i}^{\left(NC\right)},p_{j}^{\left(NC\right)}\right]=i\bar{\theta}_{ij},\,\left[x_{i}^{\left(NC\right)},p_{j}^{\left(NC\right)}\right]=i\hbar\delta_{ij},\label{eq:1}
\end{equation}
where $\tilde{\theta}_{ij}$ and $i\bar{\theta}_{ij}$ are an antisymmetric
tensor of space dimension. In order to obtain a theory which preserve
the unitary and causal, we choose $\tilde{\theta}_{0j}=0$ which implies
that the time remains as a parameter and the non-commutativity affects
only the physical space. By replacing the normal product with star
product, the Dirac equation in commuting space will change into the
Dirac equation in NC space. 
\begin{equation}
\hat{H}_{D}\left(p,x\right)\star\psi_{D}\left(x\right)=E\psi_{D}\left(x\right),\label{eq:2}
\end{equation}
where the $\star-$product Moyal between two functions is defined
by 
\begin{equation}
\left(f\star g\right)\left(x\right)=exp\left[\frac{i}{2}\tilde{\theta}_{ab}\partial_{x_{a}}\partial_{x_{b}}\right]f\left(x\right)g\left(y\right)\left|_{x=y}.\right.\label{eq:3}
\end{equation}
Instead of solving the NC Dirac equation by using the star product
procedure, we use Bopp's shift method, that is, we replace the star
product by the usual product by making a Bopp's shift 
\begin{equation}
x_{i}^{\left(NC\right)}=x_{i}-\frac{1}{2\hbar}\tilde{\theta}_{ij}p_{j},\, p_{i}^{\left(NC\right)}=p_{i}+\frac{1}{2\hbar}\bar{\theta}_{ij}x_{j}.\label{eq:4}
\end{equation}
So, in the two dimensional non-commutative phase-space, Eq. (\ref{eq:4})
becomes 
\begin{equation}
x^{\left(NC\right)}=x-\frac{\tilde{\theta}}{2\hbar}p_{y},\, y^{\left(NC\right)}=y+\frac{\tilde{\theta}}{2\hbar}p_{x},\, p_{x}^{\left(NC\right)}=p_{x}+\frac{\bar{\theta}}{2\hbar}y,\, p_{y}^{\left(NC\right)}=p_{y}-\frac{\bar{\theta}}{2\hbar}x.\label{eq:5}
\end{equation}
In this case, the two-dimensional Dirac oscillator equation, in commutative
space, which is written by
\[
\left\{ c\alpha_{x}\left(p_{x}-im_{0}\omega\tilde{\beta}x\right)+c\alpha_{y}\left(p_{y}-im_{0}\omega\tilde{\beta}y\right)+\tilde{\beta}m_{0}c^{2}\right\} \psi_{D}=E\psi_{D},
\]
\[
\]
 is modified and transformed into 
\begin{equation}
\left\{ c\alpha_{x}\left(p_{x}^{\left(NC\right)}-im_{0}\omega\tilde{\beta}x^{\left(NC\right)}\right)+c\alpha_{y}\left(p_{y}^{\left(NC\right)}-im_{0}\omega\tilde{\beta}y^{\left(NC\right)}\right)+\tilde{\beta}m_{0}c^{2}\right\} \psi_{D}=E_{NC}\psi_{D}.\label{eq:6}
\end{equation}
Using the following representation of Dirac matrices: 
\begin{equation}
\alpha_{x}=\sigma_{x}=\left(\begin{array}{cc}
0 & 1\\
1 & 0
\end{array}\right),\,\alpha_{y}=\sigma_{y}=\left(\begin{array}{cc}
0 & -i\\
i & 0
\end{array}\right),\,\tilde{\beta}=\left(\begin{array}{cc}
1 & 0\\
0 & -1
\end{array}\right),\label{eq:7}
\end{equation}
and with $\psi_{D}=\left(\begin{array}{cc}
\psi_{1} & \psi_{2}\end{array}\right)^{T}$, equation (\ref{eq:6}) becomes 
\begin{equation}
\left(\begin{array}{cc}
m_{0}c^{2} & cp_{-}\\
cp_{+} & -m_{0}c^{2}
\end{array}\right)\left(\begin{array}{c}
\psi_{1}\\
\psi_{2}
\end{array}\right)=E_{NC}\left(\begin{array}{c}
\psi_{1}\\
\psi_{2}
\end{array}\right),\label{eq:8}
\end{equation}
or 
\begin{equation}
m_{0}c^{2}\psi_{1}+cp_{-}\psi_{2}=E\psi_{1},\label{eq:9}
\end{equation}
\begin{equation}
cp_{+}\psi_{1}-m_{0}c^{2}\psi_{2}=E\psi_{2},\label{eq:10}
\end{equation}
where 
\begin{equation}
p_{-}=p_{x}^{\left(NC\right)}-ip_{y}^{\left(NC\right)}+im_{0}\omega\left(x^{\left(NC\right)}-iy^{\left(NC\right)}\right)=\varrho_{1}\left(p_{x}-ip_{y}\right)+im_{0}\omega\varrho_{2}\left(x-iy\right),\label{eq:11}
\end{equation}
\begin{equation}
p_{+}=p_{x}^{\left(NC\right)}+ip_{y}^{\left(NC\right)}-im_{0}\omega\left(x^{\left(NC\right)}+iy^{\left(NC\right)}\right)=\varrho_{1}\left(p_{x}+ip_{y}\right)-im_{0}\omega\varrho_{2}\left(x+iy\right),\label{eq:12}
\end{equation}
and where 
\begin{equation}
\varrho_{1}=1+\frac{m_{0}\omega}{2\hbar}\tilde{\theta},\,\varrho_{2}=1+\frac{\bar{\theta}}{2m_{0}\omega\hbar}.\label{eq:13}
\end{equation}
From equations (\ref{eq:9}) and (\ref{eq:10}), we have 
\begin{equation}
\left\{ c^{2}p_{-}p_{+}-\left(E^{2}-m_{0}^{2}c^{4}\right)\right\} \psi_{1}=0.\label{eq:14}
\end{equation}
Now, in order to solve the last equation, and for the sake of simplicity,
we bring the problem into the momentum space.

Recalling that 
\begin{equation}
x=i\hbar\frac{\partial}{\partial p_{x}},\, y=i\hbar\frac{\partial}{\partial p_{x}},\,\hat{p}_{x}=p_{x},\,\hat{p}_{y}=p_{y},\label{eq:15}
\end{equation}
and passing onto polar coordinates with the following definition \cite{25}
\begin{equation}
p_{x}=p\cos\theta,\, p_{y}=p\sin\theta,\, p^{2}=p_{x}^{2}+p_{y}^{2},\label{eq:16}
\end{equation}
\begin{equation}
\hat{x}=i\hbar\frac{\partial}{\partial p_{x}}=i\hbar\left(\cos\theta\frac{\partial}{\partial p}-\frac{\sin\theta}{p}\frac{\partial}{\partial\theta}\right),\label{eq:17}
\end{equation}
\begin{equation}
\hat{y}=i\hbar\frac{\partial}{\partial p_{y}}=i\hbar\left(\sin\theta\frac{\partial}{\partial p}+\frac{\cos\theta}{p}\frac{\partial}{\partial\theta}\right),\label{eq:18}
\end{equation}
Eqs. (\ref{eq:11}) and (\ref{eq:12}) transform into 
\begin{equation}
p_{-}=e^{-i\theta}\left\{ \varrho_{1}p-\lambda\left(\frac{\partial}{\partial p}-\frac{i}{p}\frac{\partial}{\partial\theta}\right)\right\} ,\label{eq:19}
\end{equation}
\begin{equation}
p_{+}=e^{i\theta}\left\{ \varrho_{1}p+\lambda\left(\frac{\partial}{\partial p}+\frac{i}{p}\frac{\partial}{\partial\theta}\right)\right\} ,\label{eq:20}
\end{equation}
where 
\begin{equation}
\lambda=\left(1+\frac{\bar{\theta}}{2m_{0}\omega\hbar}\right)m_{0}\hbar\omega.\label{eq:20.1}
\end{equation}
With the aid of these expressions, the $p_{-}p_{+}$term, appears
in the Eq. (\ref{eq:14}), can be written by 
\begin{equation}
p_{-}p_{+}=\varrho_{1}^{2}p^{2}-2\varrho_{1}\lambda-\lambda^{2}\frac{\partial^{2}}{\partial p^{2}}-\frac{\lambda^{2}}{p^{2}}\frac{\partial^{2}}{\partial\theta^{2}}-\frac{\lambda^{2}}{p}\frac{\partial}{\partial p}+2i\lambda\varrho_{1}\frac{\partial}{\partial\theta}.\label{eq:21}
\end{equation}
So, Eq. (\ref{eq:14}) becomes 
\begin{equation}
\left\{ \varrho_{1}^{2}p^{2}-\lambda^{2}\left(\frac{\partial^{2}}{\partial p^{2}}+\frac{1}{p}\frac{\partial}{\partial p}+\frac{1}{p^{2}}\frac{\partial^{2}}{\partial\theta^{2}}\right)+2i\lambda\varrho_{1}\frac{\partial}{\partial\theta}-2\lambda\varrho_{1}-\zeta\right\} \psi_{1}=0,\label{eq:22}
\end{equation}
with 
\begin{equation}
\zeta=\frac{E^{2}-m_{0}^{2}c^{4}}{c^{2}}.\label{eq:23}
\end{equation}
With the help of the following relation\cite{35} 
\begin{equation}
\psi_{1}\left(p,\theta\right)=f\left(p\right)e^{im\theta},\label{eq:24}
\end{equation}
Eq. (\ref{eq:22}) is modified and transforms into 
\begin{equation}
\left(\frac{d^{2}f\left(p\right)}{dp^{2}}+\frac{1}{p}\frac{df\left(p\right)}{dp}-\frac{m^{2}}{p^{2}}f\left(p\right)\right)+\left(\kappa^{2}-k^{2}p^{2}\right)f\left(p\right)=0,\label{eq:25}
\end{equation}
with 
\begin{equation}
\kappa^{2}=\frac{2\lambda\varrho_{1}\left(m+1\right)+\zeta}{\lambda^{2}},\, k^{2}=\frac{\varrho_{1}^{2}}{\lambda^{2}}.\label{eq:26}
\end{equation}
Putting that 
\begin{equation}
f\left(p\right)=p^{m}e^{-\frac{k^{2}}{2}p^{2}}F\left(p\right),\label{eq:27}
\end{equation}
then, the differential equation 
\begin{equation}
F^{''}+\left(\frac{2m+1}{p}-2kp\right)F^{'}-\left[2k\left(m+1\right)-\kappa^{2}\right]F=0,\label{eq:28}
\end{equation}
is obtained for $F\left(p\right)$ which by using, instead of $p$,
the variable $xt=kp^{2}$, is transformed into the Kummer equation
\begin{equation}
t\frac{d^{2}F}{dt^{2}}+\left\{ m+1-t\right\} \frac{dF}{dt}-\frac{1}{2}\left\{ m+1-\frac{\kappa^{2}}{4k}\right\} F=0,\label{eq:29}
\end{equation}
whose solution is the confluent series $_{1}F_{1}\left(a;m+1;t\right)$,
with 
\begin{equation}
a=\frac{1}{2}\left(m+1\right)-\frac{\kappa^{2}}{4k}.\label{eq:30}
\end{equation}
The confluent series becomes a polynomial if and only if $a=-n,\,\left(n=0,1,2,\right)$.

According\cite{35}we have 
\begin{equation}
\psi_{1}\left(p,\theta\right)=C_{n,m}p^{m}e^{-\frac{k}{2}p^{2}}\,_{1}F_{1}\left(-n;\left|m\right|+1;kp^{2}\right)e^{im\theta},\label{eq:31}
\end{equation}
\begin{equation}
E_{n}=\pm m_{0}c^{2}\sqrt{1+4\left(1+\frac{m_{0}\omega}{2\hbar}\tilde{\theta}\right)\left(1+\frac{\bar{\theta}}{2m_{0}\omega\hbar}\right)n}.\label{eq:32}
\end{equation}
The total associated wave function is 
\[
\psi_{n,m}\left(p,\theta\right)=\begin{pmatrix}1\\
\frac{cp_{+}}{E+m_{0}c^{2}}
\end{pmatrix}\psi_{1}.
\]
Now, in the presence of an external magnetic field, Eq. (\ref{eq:6})
is transformed into{\footnotesize{} 
\begin{equation}
\left\{ c\alpha_{x}\left[\left(p_{x}^{\left(NC\right)}+\frac{eBy^{\left(NC\right)}}{2c}\right)-im_{0}\omega\tilde{\beta}x^{\left(NC\right)}\right]+c\alpha_{y}\left[\left(p_{y}^{\left(NC\right)}-\frac{eBx^{\left(NC\right)}}{2c}\right)-im_{0}\omega\tilde{\beta}y^{\left(NC\right)}\right]+\tilde{\beta}m_{0}c^{2}\right\} \psi_{D}=\epsilon\psi_{D}.\label{eq:33}
\end{equation}
}Here the potential vectors is chosen as 
\begin{equation}
\vec{A}=\left(\begin{array}{ccc}
-\frac{By^{\left(NC\right)}}{2} & \frac{Bx^{\left(NC\right)}}{2} & 0\end{array}\right),\label{eq:34}
\end{equation}
and the Eq. (\ref{eq:33}) can be cast into a detail form as follows:
\begin{equation}
\left(\begin{array}{cc}
m_{0}c^{2} & c\tilde{p}_{-}\\
c\tilde{p}_{+} & -m_{0}c^{2}
\end{array}\right)\left(\begin{array}{c}
\tilde{\psi}_{1}\\
\tilde{\psi}_{2}
\end{array}\right)=\epsilon\left(\begin{array}{c}
\tilde{\psi}_{1}\\
\tilde{\psi}_{2}
\end{array}\right).\label{eq:35}
\end{equation}
with 
\begin{equation}
\tilde{p}_{-}=\left(p_{x}^{\left(NC\right)}+\frac{eBy^{\left(NC\right)}}{2c}\right)+im_{0}\omega x^{\left(NC\right)}-i(p_{y}^{\left(NC\right)}-\frac{eBx^{\left(NC\right)}}{2c})+m_{0}\omega y^{\left(NC\right)},\label{eq:36}
\end{equation}
\begin{equation}
\tilde{p}_{+}=\left(p_{x}^{\left(NC\right)}+\frac{eBy^{\left(NC\right)}}{2c}\right)-im_{0}\omega x^{\left(NC\right)}+i(p_{y}^{\left(NC\right)}-\frac{eBx^{\left(NC\right)}}{2c})+m_{0}\omega y^{\left(NC\right)}\label{eq:37}
\end{equation}
or 
\begin{equation}
\tilde{p}_{-}=p_{x}^{\left(NC\right)}-ip_{y}^{\left(NC\right)}+im_{0}\tilde{\omega}\left(x^{\left(NC\right)}-iy^{\left(NC\right)}\right),\label{eq:38}
\end{equation}
\begin{equation}
\tilde{p}_{+}=p_{x}^{\left(NC\right)}+ip_{y}^{\left(NC\right)}-im_{0}\tilde{\omega}\left(x^{\left(NC\right)}+iy^{\left(NC\right)}\right),\label{eq:39}
\end{equation}
where 
\begin{equation}
\tilde{\omega}=\omega-\frac{\omega_{c}}{2},\,\omega_{c}=\frac{\arrowvert e\arrowvert B}{m_{0}c}.\label{eq:40}
\end{equation}
is a cyclotron frequency.

Thus, the $\left(2+1\right)$-dimensional Dirac oscillator in a magnetic
field is mapped onto the same with reduced angular frequency $\tilde{\omega}$
in absence of magnetic field. Hence, the only role of a magnetic field
consists in reducing the angular frequency, and the entire dynamics
remains unchanged.

Using the mapping defined by (\ref{eq:5}), the systems of equations
become 
\begin{equation}
\tilde{p}_{-}=\varrho_{1}\left(p_{x}-ip_{y}\right)+im_{0}\tilde{\omega}\varrho_{2}\left(x-iy\right)\label{eq:41}
\end{equation}
\begin{equation}
\tilde{p}_{+}=\varrho_{1}\left(p_{x}+ip_{y}\right)-im_{0}\tilde{\omega}\varrho_{2}\left(x+iy\right).\label{eq:42}
\end{equation}
By the same way used above, we obtain 
\begin{equation}
\tilde{\psi}_{1}\left(p,\theta\right)=\tilde{C}_{n,m}p^{m}e^{-\frac{k}{2}p^{2}}\,_{1}F_{1}\left(-n;\left|m\right|+1;kp^{2}\right)e^{im\theta},\label{eq:43}
\end{equation}
\begin{equation}
\epsilon_{n}=\pm m_{0}c^{2}\sqrt{1+4\left(1+\frac{m_{0}\tilde{\omega}}{2\hbar}\tilde{\theta}\right)\left(1+\frac{\bar{\theta}}{2m_{0}\tilde{\omega}\hbar}\right)n}.\label{eq:44}
\end{equation}
The last equation concerning the eigenvalue is in a good agreement
with that obtained in the literature (see Ref \cite{6} and references
therein).

The corresponding total eigenfunction is given by 
\begin{equation}
\psi_{n,m}\left(p,\theta\right)=\begin{pmatrix}1\\
\frac{c\tilde{p}_{+}}{\epsilon+m_{0}c^{2}}
\end{pmatrix}\tilde{\psi}_{1}.\label{eq:40-1}
\end{equation}
\textcolor{black}{In this section, we have study the solutions of
the two dimensional Dirac oscillator with or without an external magnetic
field by using the same way described in \cite{25}: the authors work
within a momentum space representation of the Heisenberg algebra,
and by an appropriate transformation, the problem is identified as
a }Kummer differential equation\textcolor{black}{{} where the solutions
are well-know. The solutions that we have found are in well-agreement
with those obtained in the literature. This agreement allow us to
extend this method by introducing the concept of the minimal length.}

\section{The Problem with a Minimal Length}

In the minimal length formalism, the Heisenberg algebra is given by
\begin{equation}
\left[\hat{x}_{i},\hat{p}_{i}\right]=i\hbar\delta_{ij}\left(1+\beta p^{2}\right),\label{eq:45}
\end{equation}
where $\beta>0$ is the minimal length parameter. A representation
of $\hat{x}_{i}$ and $\hat{p}_{i}$ which satisfies Eq. (\ref{eq:45}),
may be taken as 
\begin{equation}
\hat{x}_{i}=i\hbar\left(1+\beta p^{2}\right)\frac{d}{dp_{i}},\,\hat{p}_{i}=p_{i},\label{eq:46}
\end{equation}
or 
\begin{equation}
\hat{x}=i\hbar\left(1+\beta p^{2}\right)\frac{d}{dp_{x}},\hat{y}=i\hbar\left(1+\beta p^{2}\right)\frac{d}{dp_{y}},\label{eq:46.1}
\end{equation}
\begin{equation}
\hat{p}_{x}=p_{x},\hat{p}_{y}=p_{y}.\label{eq:46.2}
\end{equation}
In this case, Eq. (\ref{eq:14}) is modified and becomes 
\begin{equation}
\left\{ c^{2}P_{-}P_{+}-\left(\varepsilon^{2}-m_{0}^{2}c^{4}\right)\right\} \psi_{1}=0,\label{eq:47}
\end{equation}
with 
\begin{equation}
P_{-}=\varrho_{1}\left(p_{x}-ip_{y}\right)-\lambda\left(1+\beta p^{2}\right)\left(\frac{\partial}{\partial p_{x}}-i\frac{\partial}{\partial p_{y}}\right),\label{eq:48}
\end{equation}
\begin{equation}
P_{+}=\varrho_{1}\left(p_{x}-ip_{y}\right)+\lambda\left(1+\beta p^{2}\right)\left(\frac{\partial}{\partial p_{x}}+i\frac{\partial}{\partial p_{y}}\right).\label{eq:49}
\end{equation}
In the polar coordinates, the equations (\ref{eq:48}) and (\ref{eq:49})
can be written as 
\begin{equation}
P_{-}=e^{-i\theta}\left\{ \varrho_{1}p-\lambda\left(1+\beta p^{2}\right)\left(\frac{\partial}{\partial p}-\frac{i}{p}\frac{\partial}{\partial\theta}\right)\right\} ,\label{eq:50}
\end{equation}
\begin{equation}
P_{+}=e^{i\theta}\left\{ \varrho_{2}p+\lambda\left(1+\beta p^{2}\right)\left(\frac{\partial}{\partial p}+\frac{i}{p}\frac{\partial}{\partial\theta}\right)\right\} .\label{eq:51}
\end{equation}
When we evaluate the $P_{-}P_{+}$term, we get

{\footnotesize{}
\begin{equation}
P_{-}P_{+}=\varrho_{1}^{2}p^{2}+2\left(1+\beta p^{2}\right)\left\{ \lambda\varrho_{1}\left(i\frac{\partial}{\partial\theta}-1\right)-\beta\lambda^{2}\left(p\frac{\partial}{\partial p}+i\frac{\partial}{\partial\theta}\right)\right\} -\lambda^{2}\left(1+\beta p^{2}\right)^{2}\left(\frac{\partial^{2}}{\partial p^{2}}+\frac{1}{p}\frac{\partial}{\partial p}+\frac{1}{p^{2}}\frac{\partial^{2}}{\partial\theta^{2}}\right),\label{eq:52}
\end{equation}
}and then we have

{\footnotesize{}
\begin{equation}
\left[\varrho_{1}^{2}p^{2}+2\left(1+\beta p^{2}\right)\left\{ \lambda\varrho_{1}\left(i\frac{\partial}{\partial\theta}-1\right)-\beta\lambda^{2}\left(p\frac{\partial}{\partial p}+i\frac{\partial}{\partial\theta}\right)\right\} -\lambda^{2}\left(1+\beta p^{2}\right)^{2}\left(\frac{\partial^{2}}{\partial p^{2}}+\frac{1}{p}\frac{\partial}{\partial p}+\frac{1}{p^{2}}\frac{\partial^{2}}{\partial\theta^{2}}\right)-\xi^{2}\right]\psi_{1}=0,\label{eq:53}
\end{equation}
}\textcolor{black}{with 
\begin{equation}
\xi^{2}=\frac{\varepsilon^{2}-m_{0}^{2}c^{4}}{c^{2}}.\label{eq:54}
\end{equation}
Putting that 
\begin{equation}
\psi_{1}=h\left(p\right)e^{im\theta},\label{eq:55}
\end{equation}
Eq. (\ref{eq:54}) reads}

\textcolor{black}{\small{}
\begin{equation}
\left[\varrho_{1}^{2}p^{2}-2\left(1+\beta p^{2}\right)\left\{ \lambda\varrho_{1}\left(m+1\right)+\beta\lambda^{2}\left(p\frac{d}{dp}-m\right)\right\} -\lambda^{2}\left(1+\beta p^{2}\right)^{2}\left(\frac{d^{2}}{dp^{2}}+\frac{1}{p}\frac{d}{dp}-\frac{m^{2}}{p^{2}}\right)-\xi^{2}\right]h\left(p\right)=0.\label{eq:56}
\end{equation}
}\textcolor{black}{This equation can be written in another form as
follows 
\begin{equation}
\left\{ -a\left(p\right)\frac{d^{2}}{dp^{2}}+b\left(p\right)\frac{d}{dp}+c\left(p\right)-\xi^{2}\right\} h\left(p\right)=0,\label{eq:57}
\end{equation}
where 
\begin{equation}
a\left(p\right)=\lambda^{2}\left(1+\beta p^{2}\right)^{2},\label{eq:58}
\end{equation}
\begin{equation}
b\left(p\right)=-2\beta\lambda^{2}\left(1+\beta p^{2}\right)p-\frac{\lambda^{2}\left(1+\beta p^{2}\right)^{2}}{p},\label{eq:59}
\end{equation}
\begin{equation}
c\left(p\right)=\varrho_{1}^{2}p^{2}-2\lambda\varrho_{1}\left(m+1\right)\left(1+\beta p^{2}\right)+2\beta\lambda^{2}m\left(1+\beta p^{2}\right)+\frac{\lambda^{2}\left(1+\beta p^{2}\right)^{2}m^{2}}{p^{2}}.\label{eq:60}
\end{equation}
The solutions of Eq.(\ref{eq:57}) can be found by using the following
transformations\cite{34} 
\begin{equation}
h\left(p\right)=\rho\left(p\right)\varphi\left(p\right),~q=\int\frac{1}{\sqrt{a\left(p\right)}}dp,\label{eq:61}
\end{equation}
with 
\begin{equation}
\rho\left(p\right)=e^{\int\chi\left(p\right)dp}.\label{eq:62}
\end{equation}
Using these transformations, we obtain a form similar of Schrödinger
differential equation: so 
\begin{equation}
\left(-\frac{d^{2}}{dq^{2}}+V\left(q\right)\right)\varphi\left(p\right)=\xi\varphi\left(p\right),\label{eq:63}
\end{equation}
where 
\begin{equation}
\chi\left(p\right)=\frac{2b+a^{'}}{4a}=-\frac{1}{2p},\label{eq:64}
\end{equation}
and 
\begin{equation}
V\left(p\right)=\varrho_{1}^{2}p^{2}-2\lambda\varrho_{1}\left(m+1\right)\left(1+\beta p^{2}\right)+2\beta\lambda^{2}m\left(1+\beta p^{2}\right)+\beta\lambda^{2}\left(1+\beta p^{2}\right)+\frac{\lambda^{2}\left(1+\beta p^{2}\right)^{2}}{p^{2}}\left(m^{2}-\frac{1}{4}\right).\label{eq:65}
\end{equation}
Using that 
\begin{equation}
p=\frac{1}{\sqrt{\beta}}\tan\left(q\lambda\sqrt{\beta}\right),\label{eq:66}
\end{equation}
we get 
\begin{equation}
V\left(p\right)=-\frac{1}{\beta}+\beta\lambda^{2}\left\{ \frac{\zeta_{1}\left(\zeta_{1}-1\right)}{\sin^{2}\left(q\lambda\sqrt{\beta}\right)}+\frac{\zeta_{2}\left(\zeta_{2}-1\right)}{\cos^{2}\left(q\lambda\sqrt{\beta}\right)}\right\} ,\label{eq:67}
\end{equation}
with 
\begin{equation}
\zeta_{1}\left(\zeta_{1}-1\right)=m^{2}-\frac{1}{4},\label{eq:68}
\end{equation}
\begin{equation}
\zeta_{2}\left(\zeta_{2}-1\right)=\left(m-\frac{\varrho_{1}}{\beta\lambda}+\frac{1}{2}\right)\left(m-\frac{\varrho_{1}}{\beta\lambda}+\frac{3}{2}\right)\label{eq:69}
\end{equation}
Thus, we have 
\begin{equation}
\left(-\frac{d^{2}}{dq^{2}}+\frac{1}{2}U_{0}\left\{ \frac{\zeta_{1}\left(\zeta_{1}-1\right)}{\sin^{2}\left(\alpha q\right)}+\frac{\zeta_{2}\left(\zeta_{2}-1\right)}{\cos^{2}\left(\alpha q\right)}\right\} \right)\varphi\left(p\right)=\bar{\xi}^{2}\varphi\left(p\right)\label{eq:70}
\end{equation}
with $\bar{\xi}^{2}=\xi^{2}+\frac{\alpha_{1}^{2}}{\beta}$ and $U_{0}=u^{2}$
with $u=\lambda\sqrt{\beta}$.}

\textcolor{black}{The last equation is the well-known Schrödinger
equation in a Poschl-Teller potential where\cite{33} 
\begin{equation}
U=\frac{1}{2}U_{0}\left\{ \frac{\zeta_{1}\left(\zeta_{1}-1\right)}{\sin^{2}\left(uq\right)}+\frac{\zeta_{2}\left(\zeta_{2}-1\right)}{\cos^{2}\left(uq\right)}\right\} .\label{eq:71}
\end{equation}
with $\zeta_{1}>1$ and $\zeta_{2}>1$. Thus, following Eqs. (\ref{eq:68})
and (\ref{eq:69}), we have 
\begin{equation}
\zeta_{1}=m\pm\frac{1}{2},\label{eq:72}
\end{equation}
\begin{equation}
\zeta_{2}=\frac{1}{2}\pm\left(m+1-\frac{\varrho_{1}}{\beta\lambda}\right).\label{eq:73}
\end{equation}
Introducing the new variable 
\begin{equation}
z=\sin^{2}\left(uq\right),\label{eq:74}
\end{equation}
the Schrödinger equation is transformed into 
\begin{equation}
z\left(1-z\right)\varphi^{''}+\left(\frac{1}{2}-z\right)\varphi^{'}+\frac{1}{4}\left\{ \frac{\bar{\xi}^{2}}{u^{2}}-\frac{\zeta_{1}\left(\zeta_{1}-1\right)}{z}-\frac{\zeta_{2}\left(\zeta_{2}-1\right)}{1-z}\right\} \varphi=0.\label{eq:75}
\end{equation}
}{\small{}Now, putting that 
\begin{equation}
\varphi=z^{\frac{\zeta_{1}}{2}}\left(1-z\right)^{\frac{\zeta_{2}}{2}}\Psi\left(z\right),\label{eq:76}
\end{equation}
}\textcolor{black}{we arrive at 
\begin{equation}
z\left(1-z\right)\Psi^{''}+\left[\left(\zeta_{1}+\frac{1}{2}\right)-z\left(\zeta_{1}+\zeta_{2}+1\right)\right]\Psi^{'}+\frac{1}{4}\left\{ \frac{\bar{\xi}^{2}}{u^{2}}-\left(\zeta_{1}+\zeta_{2}\right)^{2}\right\} \Psi=0.\label{eq:77}
\end{equation}
The general solutions of this equation are\cite{35,36} 
\begin{equation}
\Psi=C_{1}\,_{2}F_{1}\left(a;b;c;z\right)+C_{2}\, z^{1-c}\,_{2}F_{1}\left(a+1-c;b+1-c;2-c;z\right),\label{eq:78}
\end{equation}
where 
\begin{equation}
a=\frac{1}{2}\left(\zeta_{1}+\zeta_{2}+\frac{\bar{\xi}}{u^{2}}\right),\, b=\frac{1}{2}\left(\zeta_{1}+\zeta_{2}-\frac{\bar{\xi}}{u^{2}}\right),\, c=\zeta_{1}+\frac{1}{2}.\label{eq:79}
\end{equation}
With the condition $a=-n$, we obtain 
\begin{equation}
\bar{\xi}^{2}=u^{2}\left(\zeta_{1}+\zeta_{2}+2n\right)^{2}.\label{eq:80}
\end{equation}
In order to obtain the energy of spectrum, it should be to note that
in the limit $\beta\rightarrow0$, the energy of spectrum should be
covert to no-GUP result. Thus, we choose 
\begin{equation}
\zeta_{1}=m+\frac{1}{2},\label{eq:81}
\end{equation}
\begin{equation}
\zeta_{2}=\frac{1}{2}-\left(m+1-\frac{\varrho_{1}}{\beta\lambda}\right).\label{eq:82}
\end{equation}
Following this, we obtain 
\begin{equation}
\epsilon_{n}=\pm\sqrt{m_{0}^{2}c^{4}+4c^{2}\left(1+\frac{m_{0}\omega}{2\hbar}\tilde{\theta}\right)\left(1+\frac{\bar{\theta}}{2m_{0}\omega\hbar}\right)n+4c^{2}\beta\left(1+\frac{\bar{\theta}}{2m_{0}\omega\hbar}\right)^{2}n^{2}},\label{eq:83}
\end{equation}
where 
\begin{equation}
\beta<\beta_{0},\,\beta_{0}=\frac{1}{m+\frac{3}{2}}\frac{\left(1+\frac{m_{0}\omega}{2\hbar}\tilde{\theta}\right)}{\left(1+\frac{\bar{\theta}}{2m_{0}\omega\hbar}\right)\hbar\omega m_{0}},\,\mbox{with }m>0.\label{eq:84}
\end{equation}
So, the non-zero minimal length is 
\begin{equation}
\triangle x_{\mbox{min}}=\hbar\sqrt{\beta}<\left(\triangle x_{\mbox{min}}\right)_{0}=\sqrt{\frac{1}{m+\frac{3}{2}}\frac{\left(1+\frac{m_{0}\omega}{2\hbar}\tilde{\theta}\right)}{\left(1+\frac{\bar{\theta}}{2m_{0}\omega\hbar}\right)}}l_{min},\label{eq:85}
\end{equation}
with $l_{min}=\sqrt{\frac{\hbar}{m_{0}\omega}}$ is the characteristic
length of the Dirac oscillator, and $\left(\triangle x_{\mbox{min}}\right)_{0}$
is the admissible length above it the physics becomes experimentally
inaccessible. We can see that the influence of the non-commutative
parameters on $\left(\triangle x_{\mbox{min}}\right)_{0}$ is very
clear. Now, expanding to first order in $\beta$ we have \cite{20}}

\textcolor{black}{\small{}
\begin{eqnarray}
{\color{black}\epsilon_{n}} & {\color{black}\simeq} & {\color{black}\pm m_{0}c^{2}\sqrt{1+\frac{4}{m_{0}^{2}c^{2}}\left(1+\frac{m_{0}\omega}{2\hbar}\tilde{\theta}\right)\left(1+\frac{\bar{\theta}}{2m_{0}\omega\hbar}\right)n}\times}\nonumber \\
{\color{black}} & {\color{black}} & {\color{black}\left(1+\frac{2\beta\left(1+\frac{\bar{\theta}}{2m_{0}\omega\hbar}\right)}{m_{0}^{2}c^{2}}\frac{n^{2}}{1+\frac{4}{m_{0}^{2}c^{2}}\left(1+\frac{m_{0}\omega}{2\hbar}\tilde{\theta}\right)\left(1+\frac{\bar{\theta}}{2m_{0}\omega\hbar}\right)n}\right)}\label{eq:90-1}
\end{eqnarray}
}{\small \par}

\textcolor{black}{The first term is the energy spectrum of the usual
two-dimensional Dirac oscillator and the second term represents the
correction due the presence of the minimal length. As mentioned by\cite{19},
we note the dependence on $n^{2}$ which is feature of hard confinement.
For a large values of $n$ we have 
\begin{equation}
\epsilon_{n}=\hbar\bar{\omega}n,\label{eq:87}
\end{equation}
which means, according that the energy continuum for large $n$ for
the Dirac oscillator without the minimal length disappears in the
presence of the minimal length, and consequently the behavior of the
DO can be described by a non-relativistic harmonic oscillator with
frequency $\bar{\omega}=\frac{2c\sqrt{\beta}}{\hbar}\left(1+\frac{\bar{\theta}}{2m_{0}\omega\hbar}\right)$.}

\textcolor{black}{According to the Eqs. (\ref{eq:79}) and (\ref{eq:81}),
we can see that the parameter $c=m+1$ is an integer: thus either
the two solutions of Eq. (\ref{eq:78}) coincide or one of the solutions
will blow up. Now, when $c$ is an integer greater than 1, which is
our case, the second solution diverges. Thus, the component $\psi_{1}$
will has the following form 
\begin{equation}
\left(\psi_{1}\right)_{n,m}\left(p,\theta,z\right)=\left(C_{1}\right)_{n,m}p^{-\frac{1}{2}}e^{im\theta}z^{\frac{\zeta_{1}}{2}}\left(1-z\right)^{\frac{\zeta_{2}}{2}}\,_{2}F_{1}\left(-n;b,\left|m\right|+1;z\right).\label{eq:86-2}
\end{equation}
Finally, the total associated eigenfunction is done by 
\begin{equation}
\psi_{n,m}\left(p,\theta,z\right)=\begin{pmatrix}1\\
\frac{cP_{+}}{\epsilon+m_{0}c^{2}}
\end{pmatrix}\psi_{1}.\label{eq:86-1-1}
\end{equation}
Now, in the presence of an uniform magnetic field, Eq. (\ref{eq:6})
is transformed into}\textcolor{black}{\footnotesize{}
\begin{equation}
\left\{ c\alpha_{x}\left[\left(p_{x}^{\left(NC\right)}+\frac{eBy^{\left(NC\right)}}{2c}\right)-im_{0}\omega{\color{black}}\tilde{\beta}x^{\left(NC\right)}\right]+c\alpha_{y}\left[\left({\color{black}}p_{y}^{\left(NC\right)}-\frac{eBx^{\left(NC\right)}}{2c}\right)-im_{0}\omega\tilde{\beta}y^{\left(NC\right)}\right]+\tilde{\beta}m_{0}c^{2}\right\} \psi_{D}=\bar{\epsilon}\psi_{D}.\label{eq:94}
\end{equation}
}\textcolor{black}{{} In this case, Eq. (\ref{eq:8}) takes the following
form 
\begin{equation}
\left(\begin{array}{cc}
m_{0}c^{2} & c\tilde{P}_{-}\\
c\tilde{P}_{+} & -m_{0}c^{2}
\end{array}\right)\left(\begin{array}{c}
\psi_{1}\\
\psi_{2}
\end{array}\right)=\bar{\epsilon}\left(\begin{array}{c}
\psi_{1}\\
\psi_{2}
\end{array}\right),\label{eq:88}
\end{equation}
with 
\begin{equation}
\tilde{P}_{-}=\left(p_{x}^{\left(NC\right)}+\frac{eBy^{\left(NC\right)}}{2c}\right)+im_{0}\omega x^{\left(NC\right)}-i(p_{y}^{\left(NC\right)}-\frac{eBx^{\left(NC\right)}}{2c})+m_{0}\omega y^{\left(NC\right)},\label{eq:89}
\end{equation}
\begin{equation}
\tilde{P}_{+}=\left(p_{x}^{\left(NC\right)}+\frac{eBy^{\left(NC\right)}}{2c}\right)-im_{0}\omega x^{\left(NC\right)}+i(p_{y}^{\left(NC\right)}-\frac{eBx^{\left(NC\right)}}{2c})+m_{0}\omega y^{\left(NC\right)},\label{eq:90}
\end{equation}
or 
\begin{equation}
\tilde{P}_{-}=\varrho_{1}\left(p_{x}-ip_{y}\right)+im_{0}\tilde{\omega}\varrho_{2}\left(x-iy\right),\label{eq:91}
\end{equation}
\begin{equation}
\tilde{P}_{+}=\varrho_{1}\left(p_{x}+ip_{y}\right)+im_{0}\tilde{\omega}\varrho_{2}\left(x+iy\right),\label{eq:92}
\end{equation}
where 
\begin{equation}
\tilde{\omega}=\omega-\frac{\omega_{c}}{2},\,\omega_{c}=\frac{\arrowvert e\arrowvert B}{m_{0}c},\label{eq:93}
\end{equation}
and where $\omega_{c}$ is a cyclotron frequency. According to the
above case, the eigensolutions are given by 
\begin{equation}
\left(\psi_{1}\right)_{n,m}\left(p,\theta,z\right)=\left(\tilde{C}_{1}\right)_{n,m}p^{-\frac{1}{2}}e^{im\theta}z^{\frac{\zeta_{1}}{2}}\left(1-z\right)^{\frac{\zeta_{2}}{2}}\,_{2}F_{1}\left(-n;b,\left|m\right|+1;z\right),\label{eq:97}
\end{equation}
\[
\bar{\epsilon}_{n}=\pm\sqrt{m_{0}^{2}c^{4}+4c^{2}\left(1+\frac{m_{0}\tilde{\omega}}{2\hbar}\tilde{\theta}\right)\left(1+\frac{\bar{\theta}}{2m_{0}\tilde{\omega}\hbar}\right)n+4c^{2}\beta\left(1+\frac{\bar{\theta}}{2m_{0}\tilde{\omega}\hbar}\right)^{2}n^{2}},
\]
where the total wave-function is done by 
\begin{equation}
\psi_{n,m}\left(p,\theta,z\right)=\begin{pmatrix}1\\
\frac{c\tilde{P}_{+}}{\bar{\epsilon}+m_{0}c^{2}}
\end{pmatrix}\psi_{1}.\label{eq:98-1}
\end{equation}
}

\section{Results and Discussions}

\textcolor{black}{Here we have obtained exact solutions of the two-dimensional
Dirac oscillator in noncommutative space with the presence of minimal
length. Firstly, by adopting the same procedure that used by Menculini
et al\cite{25}, we have solved the problem only in the case of noncommutative
space. The results found are in well agreement with those obtained
in the literature. After that, we have introduced the minimal length
in the the problem in question. This introduction has been make as
follows: (i) we write the coordinates of the noncommutative space
with those in commutative space by using the Bipp shift approximation,
and (ii) then we introduce the minimal length in our equation. By
these, the problem in question is identified with a Poschl-Teller
potential. Also, when $\theta$, and $\bar{\theta}$ tend to zero,
we recovert exactly the same results of \cite{37}.}

\textcolor{black}{Finally, let us note that the non-relativistic harmonic
oscillator is used as a model for describing the quark\textquoteright s
confinement in mesons and baryons, while the Dirac oscillator is expected
to give a good description of the confinement in heavy quark systems.
Quimby and Strange suggested that the two-dimensional Dirac oscillator
model can be describe some properties of electrons in graphene. This
model explains the origin of the left-handed chirality observed for
charge carriers in monolayer and bilayer graphene. They have shown
that the change of the strength of a magnetic field leads to the existence
of a quantum phase transition in the chirality of the systems. In
addition, in a recent paper, it has been shown that we can modulate
the system of graphene under a magnetic field with a model based on
a Dirac oscillator. With this, the author has determine all thermodynamic
properties of this system by using the thermal zeta function \cite{38}.}

\textcolor{black}{In our case, a possible application is the determination
of the upper limit of the length in comparison with the data found
experimentally for the case of graphene: this idea has been used by
Menculini et al \cite{25} in order to obtain an upper bound on the
minimal length appearing in the framework of generalized uncertainly
principle. }

\section{Conclusion}

In this paper, we have exactly solved the Dirac oscillator in two
dimensions in the presence of an external magnetic field in the framework
of relativistic quantum mechanics with minimal length and in the noncommutative
phase-space. Firstly, the eigensolutions of the problem in question
are obtained in noncommutative space. Then, we extend our study in
the presence of a minimal length. The energy levels, for both cases,
show a dependence on $n^{2}$ in the presence of the minimal length
which described a hard confinement. For the large values of $n$,
our DO become like a non-relativistic harmonic oscillator. the dependence
of the non-zero minimum length on the noncommutativite parameters
is very clear. In the limit where $\beta\rightarrow0$, \textcolor{black}{and
where $\theta$, and $\bar{\theta}$ tend to zero, }we recover the
results obtained in the literature

\end{document}